# Entropic thermodynamics of nonlinear photonic chain networks

Fan O. Wu[1,4], Pawel S. Jung[1,2,4], Midya Parto[1], Mercedeh Khajavikhan[3], Demetrios N. Christodoulides[1,*]

[1]CREOL, College of Optics and Photonics, University of Central Florida, Orlando, Florida 32816–2700, USA.

[2]Faculty of Physics, Warsaw University of Technology, Koszykowa 75, 00-662 Warsaw, Poland

[3]Department of Electrical and Computer Engineering, University of Southern California, Los Angeles, California 9007, USA

[4]These authors contributed equally

*demetri@creol.ucf.edu

Abstract

The complex nonlinear behaviors of heavily multimode lightwave structures have been recently the focus of considerable attention. Here we develop an optical thermodynamic approach capable of describing the thermalization dynamics in large scale nonlinear photonic chain networks – a problem that has remained unresolved so far. A Sackur-Tetrode equation is obtained that explicitly provides the entropy of these systems in terms of all the extensive variables involved. This entropic description leads to a self-consistent set of equations of state from where the optical temperature and chemical potential of the photon gas can be uniquely determined. Processes like isentropic expansion/compression, Joule expansion, as well as aspects associated with beam cleaning/cooling and thermal conduction effects in such chain networks are discussed. Our results open new vistas through which one can describe in an effortless manner the exceedingly complex dynamics of highly multimoded nonlinear bosonic systems that are nowadays of importance in areas ranging from high-power photonic settings to cold atom condensates and optomechanics.

In recent years, there has been considerable interest in understanding the rich and complex dynamics of large-scale nonlinear multimoded optical systems[1-5]. To a great extent, what has fueled this effort is the ever increasing quest for high information capacity photonic networks[6-10] and high-power optical sources[11]. Similarly, the nonlinear physics of high-Q multimode microresonators has been intensively investigated in a number of areas, like comb-generation and optomechanics[12-16]. Yet, what kept us from fully expanding these classical settings, is the sheer complexity behind the nonlinear interactions taking place in such vastly many-moded environments. To appreciate this difficulty, consider for example a multimode optical fiber supporting thousands of modes. In this case, to correctly account for all the nonlinear processes unfolding in this structure, one has to first estimate at least a trillion or so four-wave mixing coefficients, before even attempting to numerically solve the evolution problem – a formidable task which is by itself computationally intensive. Clearly, to overcome these hurdles, an altogether new approach has to be deployed that can effectively deal with "many-body" configurations. In essence, this calls for the development of a theoretical framework, akin to that of statistical mechanics, that has so far allowed us to understand the macroscopic properties of matter, in spite of the fact that on most occasions one has to deal with a great multitude of atoms or molecules – most often exceeding Avogadro's number[17].

Here, by means of an optical Sackur-Tetrode equation, we investigate the entropic response of large-scale weakly nonlinear photonic chain networks. These classical networks may for example involve a large number of evanescently coupled waveguide lattices[18] that exchange power in both crystalline (Fig. 1a) and other more complex arrangements (Fig. 1b), or could be comprised of arrays of optical resonators (CROWs)[19] that allow for energy transfer in time – used as building blocks as shown in Fig. 1c. While in this work, we will be primarily dealing with coupled waveguide topologies in order to exemplify these notions, a similar discussion holds for heavily multimode chain cavity configurations and other bosonic systems[20]. In this case, the underlying Sackur-Tetrode equation can be expressed as a function of the electrodynamic momentum flow $U$, the number of available modes $M$, and the input optical power $\mathcal{P}$. This analytical formalism developed here is in turn used to study and predict several distinct phenomena that could be of practical importance. These include for example, the prospect for beam cooling (leading to an improvement in beam quality) as well as Joule expansion and aspects associated with the photo-capacity and thermal conductivity of such chain networks.

We begin our discussion by considering a lossless, nonlinear array network of $M$ single-mode elements that are coupled via nearest-neighbor interactions. In this nonlinear optical chain, the equation of motion are given by $ida_m/d\xi + \kappa a_{m-1} + \kappa a_{m+1} + |a_m|^2 a_m = 0$, where $a_m$ represents the local mode amplitude at site $m$ and $\kappa$ stands for the coupling coefficient between two successive elements. These evolution equations can be obtained from the classical Hamiltonian $\langle H_T \rangle = \sum_{m,n} \kappa(a_{m-1}a_n^* + a_{m+1}a_n^*) + (1/2)\sum_m |a_m|^4$. The linear eigenvalues $\varepsilon_j$ corresponding to the optical supermodes $|\phi_j\rangle$ of the arrangement shown in Fig. 1a are given by $\varepsilon_j = 2\kappa \cos[\pi j/(M+1)]$ (ref. [18,19]). In general, the evolution of this weakly nonlinear system can be described in terms of its supermodes, i.e., $|\psi\rangle = \sum_{j=1}^{M} c_j(\xi)|\phi_j\rangle$ where the square of the moduli ($|c_j|^2$) of the complex coefficients $c_j$ denotes modal occupancies. In this case, the norm in this conservative system $\mathcal{P} = \sum_{j=1}^{M}|c_j|^2$ is preserved, thus implying power/photon conservation in these coupled waveguide/cavity arrangements. In addition, in the weakly nonlinear regime, the expectation value of the total Hamiltonian $\langle H_T \rangle = \langle H_L \rangle + \langle H_{NL} \rangle$ is dominated by its linear component $\langle H_L \rangle$ and as a result the internal energy $U = -\langle\psi|H_L|\psi\rangle = -\sum_{j=1}^{M}\varepsilon_j|c_j|^2$ remains invariant during evolution, thus establishing the second conservation law associated with this class of structures. In fact, in multimode waveguide settings, the conserved internal energy $U$ so happens to be the longitudinal electrodynamic momentum flow in these systems[21]. In this respect, the two normalized constants of the motion $\mathcal{P}$ and $U$ are uniquely determined by the initial excitation amplitudes $c_{j0}$, since $\mathcal{P} = \sum_{j=1}^{M}|c_{j0}|^2$ and $U = -\sum_{j=1}^{M}\varepsilon_j|c_{j0}|^2$. At this point, it is important to stress that the sole role of the weak nonlinearity involved ($\chi^{(2)}, \chi^{(3)}$, etc.) is to chaotically reshuffle the optical power among modes through multi-wave mixing processes[22] – thus causing the complex amplitudes $c_j$ to vary randomly during propagation. This in turn, allows the system to ergodically explore in a fair manner, all its accessible microstates (in its phase space) that lie on the constant energy ($U$) and power ($\mathcal{P}$) manifolds. It is this ergodicity that provides the foundation for establishing an entropic thermodynamic theory for the aforementioned multimode chain networks. Moreover, in heavily multimode arrangements, the thermodynamically extensive variables ($U, M, \mathcal{P}$) are related to each other through the optical temperature $T$ and chemical potential $\mu$ associated with the system. This relation is given through a global equation of state, given by $U - \mu\mathcal{P} = MT$ that explicitly involves the total number of modes $M$ [23,24]. In general, any system is expected to reach thermal equilibrium by maximizing its entropy. In this case, the thermalized average power levels conveyed by each mode are found to obey a Rayleigh-Jeans distribution[25-28], i.e. $|c_j|^2 = -T/(\varepsilon_j + \mu)$.

Based on the above premises, one can show that the relative entropy associated with a photonic "monoatomic" chain network (Fig. 1) is given by the following equation[24], i.e.,

$$S = S(U, M, \mathcal{P}) = M \ln\left(\frac{4\kappa^2\mathcal{P}^2 - U^2}{4M\kappa^2\mathcal{P}}\right). \quad (1)$$

Equation (1) is analogous to the entropic Sackur-Tetrode equation – developed more than a century ago (based on an appropriate quantization of the phase space) in order to correctly account for the properties of monoatomic gases[17,29]. Like its counterpart in statistical mechanics, the optical entropy of these nonlinear monoatomic chains is extensive with respect to the three other extensive variables $(U, M, \mathcal{P})$. In other words, if we let $(U, M, \mathcal{P}) \to (\lambda U, \lambda M, \lambda \mathcal{P})$, then from Eq. (1), one directly obtains $S \to \lambda S$. We note that this extensivity in entropy is crucial in developing a self-consistent thermodynamic formulation – free of Gibbs paradoxes. On the other hand, the optical entropy given above is by nature different from that of an ideal monoatomic gas in standard thermodynamics. This is apparent, since the prefactor in Eq. (1) is now given by the number of modes $M$ ("optical volume") as opposed to the optical power $\mathcal{P}$ which here plays the role of $N$ – the number of gas particles. Figure 2a depicts the entropy of a chain system as a function of $(U, \mathcal{P})$ as obtained from Eq. (1), when $M = 100$. The three pertinent equations of state can in turn be derived from (1) by employing the fundamental equation of thermodynamics[24], i.e.,

$$\frac{1}{T} = \frac{\partial S}{\partial U} = \frac{2MU}{U^2 - 4\kappa^2\mathcal{P}^2} \quad (2a)$$

$$\frac{\mu}{T} = -\frac{\partial S}{\partial \mathcal{P}} = \frac{M}{\mathcal{P}} + \frac{8M\kappa^2\mathcal{P}}{U^2 - 4\kappa^2\mathcal{P}^2} \quad (2b)$$

$$\frac{\hat{p}}{T} = \frac{\partial S}{\partial M} = \ln\left(\frac{4\kappa^2\mathcal{P}^2 - U^2}{4M\kappa^2\mathcal{P}}\right) - 1 \quad (2c)$$

Equations (2), relate the three intensive thermodynamic variables $(T, \mu, \hat{p})$ to their respective conjugate quantities $(U, \mathcal{P}, M)$. As in statistical mechanics, in this nonlinear multimoded setting, the optical temperature $T = (U^2 - 4\kappa^2\mathcal{P}^2)/(2MU)$ represents a thermodynamic force responsible for energy transfer $\Delta U$ from a hot to a colder subsystem whereas the chemical potential $\mu$ dictates the power exchange $\Delta\mathcal{P}$ – all aiming to maximize the entropy in accord with the second law of thermodynamics. Finally, $\hat{p}$ represents the relative optical pressure in this configuration. From Eq. (2a), one can readily deduce that as the system attains its lowest possible internal energy $U_{min} \to -2\kappa\mathcal{P}$ (as imposed by the band structure of the monoatomic lattice), its temperature tends to zero ($T \to 0^+$) and as a result the photon gas becomes a condensate that exclusively occupies the ground state of the lattice. Conversely, when the internal energy is maximized $U_{max} \to 2\kappa\mathcal{P}$, the opposite is true (i.e., $T \to 0^-$), and hence the highest order mode is now solely occupied (anti-condensate)[30]. The first equation of state (Eq. (2a)) directly indicates that the temperature of these lattices is positive for $-2\kappa\mathcal{P} < U < 0$ whereas is negative if the internal energy lies in the domain $0 < U < 2\kappa\mathcal{P}$. On the other hand, when $U \to 0^{\mp}$, the temperature tends to $T \to \pm\infty$ (as schematically shown in the inset of Fig 2b), and as a result, equipartition of power takes place among modes, i.e., $|c_j|^2 = -T/\mu = \mathcal{P}/M$. In general, negative temperatures correspond to bodies "hotter" than hot[17] – a relation that regulates the direction of the energy flow $\Delta U$. While for positive temperatures the lower group of modes is mostly occupied (since $|c_j|^2 = -T/(\varepsilon_j + \mu)$), the opposite is true for negative temperatures where the higher-order modes are populated (Fig 2b). We note that, the set of the three equations in (2) is complete, and is formally equivalent to the global equation of state ($U - \mu\mathcal{P} = MT$) as well as to the corresponding Euler equation ($S = (U - \mu\mathcal{P} + \hat{p}M)/T$) – a relation that again reaffirms the extensivity of the entropy itself[24]. Meanwhile, Fig. 2c depicts energy-temperature curves ($U - T$) at various power levels, that as we will see dictate the photo-capacity of these photonic chain networks.

The above results can be used to predict the outcome of more complex processes like, for example, that associated with Joule expansion of the photon gas in such nonlinear heavily multimoded environments. This prospect is shown schematically in Fig. 3, where as an example, $\chi^{(3)}$ nonlinear array supporting $M$ modes suddenly expands to four times its size ($M \to 4M$), while all the lattice parameters are kept the same. In this scenario, both the internal energy $U$ and power $\mathcal{P}$ remain constant during this abrupt transition. From Eqs. (2a-b), one can quickly deduce that after Joule expansion, the temperature is reduced to one fourth of its original value ($T \to T/4$), while the chemical potential $\mu$ is entirely unaffected. This photonic response is in stark contrast to the Joule expansion behavior expected from ideal monoatomic gases where the temperature is constant whereas the chemical potential substantially changes. On the other hand, even in this case, the absolute entropy of the system always increases in response to this irreversible expansion[24]. In a similar vein, isentropic processes ($S = const$) can also be examined. These effects can be readily produced if, for example, a Kerr nonlinear multimode array adiabatically expands/contracts (in which case the mode occupancies $|c_j|^2$ are unaltered) while the individual single-mode elements remain the same (only the coupling coefficient $\kappa$ adiabatically changes). Since $M$ and $\mathcal{P}$ are invariant, an isentropic transition can only occur provided that $U/\kappa = const$ (ref. [24]). This last relation can be justified given that $U = -\sum_{j=1}^{M} \varepsilon_j |c_j|^2$ and $\varepsilon_j \propto \kappa$. From the first two equations of state (Eqs. (2a-b)), one can quickly conclude that $U/T = const$ and $\mu/T = const$. These latter isentropic relations, applicable for the photon gas in multimode chain networks, are to some extent reminiscent of their counterparts in the theory of monatomic gases, i.e., $pV^{5/3} = const$ and $TV^{2/3} = const$ (ref. [17]). Nevertheless, there is a fundamental difference between these two physical settings, in the sense that in the former case there is a physical change in the lattice structure (and hence in the width of the density of states) while in the latter the gaseous substance stays the same.

The analytical formalism (Eqs. (2)) developed here, can now be utilized to predict the response of such nonlinear chain systems in settings of practical interest like those allowing for beam self-cleaning[31-34] and cooling. Figure 4a schematically shows such a nonlinear multi-core optical arrangement involving $M = 30$ sites. In this configuration, the power ($\mathcal{P} = 2$) is injected in the $\hat{y}$ polarization and is uniformly distributed (with random phases) between the supermodes having eigenvalues $\varepsilon_j$ in the range of $-0.5\kappa \leq \varepsilon_j \leq 1.2\kappa$, in which case $U = -0.77$, as shown in Fig. 4b. For these initial conditions, Eqs. 2 predict the following equilibrium parameters $T = 0.33, \mu = -5.38$. While at the input ($z = 0$) the intensity distribution displays considerable disorder (Fig. 4c), this speckle pattern disappears (Fig. 4c) after the system attains thermal equilibrium at the output ($z = L$). What facilitates this "beam self-cleaning" effect has to do with the fact that during thermalization, the power is reshuffled in such a way so as to favor the lower order modes (for $T > 0$) in the Rayleigh-Jeans distribution (Fig.4b) – a process that removes the initial speckle in the beam. Yet, while the output beam at ($z = L$) seems to improve in the near-field, its multimode far-field is still highly divergent and thus is of poor quality as shown in Fig. 4d. Interestingly, the thermodynamic formulation developed here suggests that this situation can be greatly improved if in this same array structure, a "cooler" beam is launched – having a perpendicular polarization $\hat{x}$ (Fig. 4e). In this latter arrangement, each waveguide site is assumed to be substantially birefringent, thus prohibiting any power exchange between the two beams. Instead, the two wavefronts only interact through cross-phase modulation[23]. In the example provided in Fig. 4e, the internal energy and power corresponding to each polarization is $U_x = -3.95, \mathcal{P}_x = 2, U_y = -0.77, \mathcal{P}_y = 2$. Based on these initial conditions, Eqs. 2 indicate that on its own, each polarization would have settled to an equilibrium temperature of $T_x = 1.7 \times 10^{-3}, T_y = 0.33$. On the other hand, once interacting together, the two beams reach the same temperature $T = 0.075$ which can be exactly predicted following the procedure of ref. [24]. In other words, the $\hat{y}$ polarized beam can be considerably cooled after exchanging energy $\Delta U$ with its $\hat{x}$ polarized counterpart. This in turn leads to a ~3-fold improvement in the far-field of the original beam (Fig. 4f).

The formalism developed here, allows one to define a photo-capacity $C_M = \partial U/\partial T$ associated with such nonlinear multimode photonic systems – a property that is in every respect analogous to that of the heat capacity pertaining to various phases of matter. Like its counterpart (heat capacity), this quantity critically depends on the density of states and hence it is characteristic of the multimode optical system itself. In this respect, it is formally possible to assign a $C_M$ capacity to any possible many-mode arrangement, irrespective of whether it is discrete or continuous. For the chain networks considered here, the photo-capacity can be explicitly obtained from Eq. (2a) and it is given by $C_M = M - |T|M^2(T^2M^2 + 4\kappa^2\mathcal{P}^2)^{-1/2}$. Figure 5a shows the $C_M$ function (blue curve) for such a chain configuration, involving $M = 200$ sites while conveying a power of $\mathcal{P} = 20$. In this case, the photo-capacity function is an even function of temperature – as one should expect from the symmetry in the density of states. The photo-capacity curves corresponding to two different types of Lieb lattices are also displayed, as an example, in Fig. 5a for the same parameters ($M = 200, \mathcal{P} = 20$). Note that the $C_M(T)$ function for the Lieb-2 lattice with cross-couplings is in this case highly asymmetric – a feature attributed to the band structure of this specific system.

We next explore the possibility for energy $\Delta U$ transport phenomena occurring when a hot and a cold optical nonlinear multimode system are linked together under non-equilibrium conditions. To some extent, these processes have much in common with heat conduction problems in solids. Figure 5b shows such an arrangement where a square multicore photonic lattice $L_1$ with $M_1 = 900$ sites is connected to a graphene-like system $L_2$ ($M_2 = 896$) via a chain bridge $L_3$ involving $M_3 = 140$ evanescently coupled waveguides. The bridge $L_3$ is appropriately designed using birefringent elements[24] so as to allow for energy exchange $\Delta U$ between the two lattices $L_1$ and $L_2$ while prohibiting any power transfer $\Delta \mathcal{P}$. System $L_1$ together with the link array $L_3$ are initially brought to thermal equilibrium where both share a temperature of $T_1 = 0.3$ when conveying a total power of $\mathcal{P}_{1+3} = 65$ at a total internal energy $U_{1+3} = -60$. Similarly, lattice $L_2$ (before is connected to the bridge) is kept at $T_2 = 0.1$ when $\mathcal{P}_2 = 40$ and $U_2 = -46$. Once $L_2$ is put in contact to the $L_3$ bridge, a non-equilibrium thermodynamical process ensues during which the entropy of the combined system starts to increase as expected from the second law of thermodynamics (Fig. 5c). During this non-equilibrium stage, internal energy $\Delta U$ starts to flow through the bridge from hot to cold ($L_1 \rightarrow L_2$). In this case, the local temperature along this bridge (obtained after projecting on the local eigenfunctions) is found to linearly decrease from $T_1$ to $T_2$ in full accord to Fick's law of heat diffusion[35], $\vec{J} = -k\,\nabla T$. This in turn allows one to express this energy transfer through an effective "photo-conduction coefficient" $K$ as $\Delta U/\Delta z = -K(T_1 - T_2)$. From our simulations, we estimate that here the photo-conduction coefficient of the $L_3$ chain is $K \approx 3.2 \times 10^{-5}$.

By harnessing concepts from statistical mechanics, we have developed an optical thermodynamic theory that explicitly provides the Sackur-Tetrode entropy associated with large nonlinear photonic chain networks. Based on these premises, the temperature and chemical potential of such systems can be obtained in closed form as a function of the initial excitation conditions. The formalism developed here is general and can be readily deployed to describe a number of thermodynamic processes, like isentropic expansion/compression, Joule expansion, as well as aspects pertaining to beam self-cleaning and cooling in such chain networks. Our results not only provide a platform to understand and predict in a systematic way the utterly complex dynamics of heavily multimoded nonlinear optical systems but could also be of practical importance in designing new classes of high-brightness multimode optical sources.

**Acknowledgements** This work was supported by the Office of Naval Research (ONR) (MURI N00014-17-1-2588), Office of Naval Research (ONR) (N00014-18-1-2347), National Science Foundation (NSF) (EECS-1711230), and Qatar National Research Fund (QNRF)(NPRP9-020-1-006). P. S. J. thanks the Polish Ministry of Science and Higher Education for Mobility Plus scholarship.

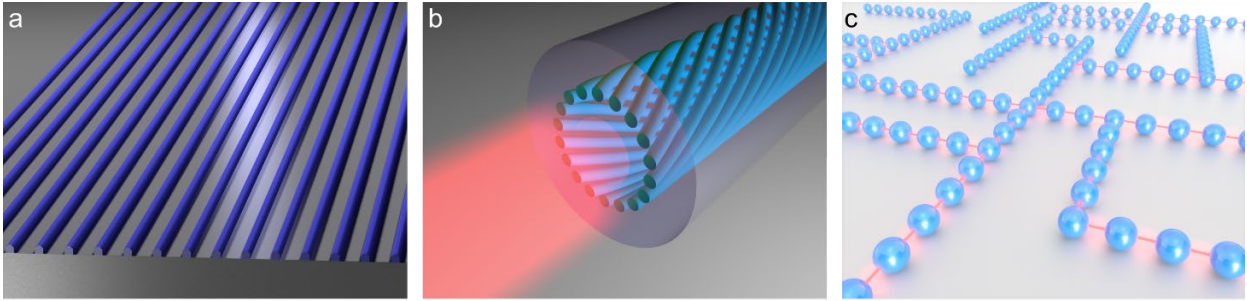

**Fig. 1 | Nonlinear multimode optical chain networks. a-c,** This class of networks may come in a variety of forms, like (**a**) a large lattice of evanescently coupled waveguide elements (**b**) a twisted topological array system and (**c**) an optical coupled-cavity lattice configuration where energy is exchanged in time.

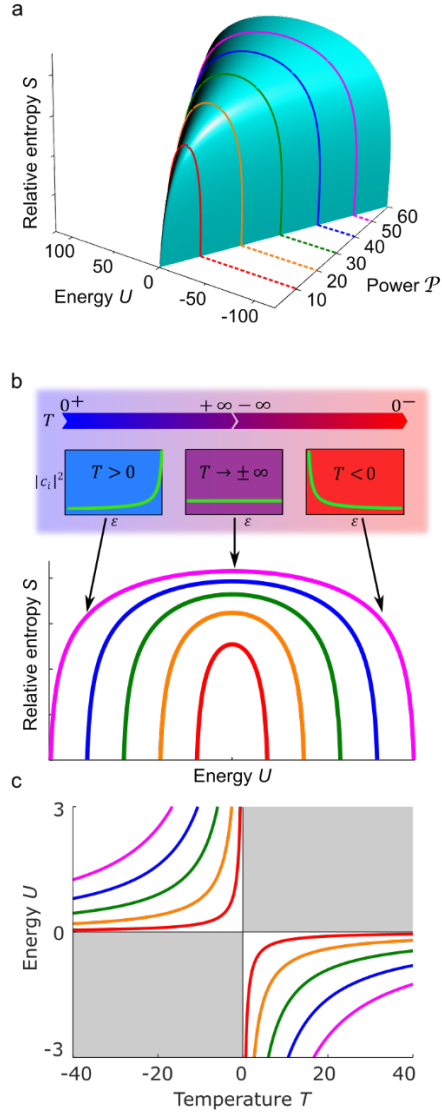

**Fig. 2 | Optical entropy and $U-T$ diagrams associated with a nonlinear multimode chain network.**
**a**, Relative entropy $S$ as a function of the internal energy $U$ and input power $\mathcal{P}$. **b**, Entropy cross-sections as a function of energy $U$ for the power levels indicated in (**a**) with same colored curves. The inset in (**b**) depicts the modal occupancies corresponding to different temperatures. While for positive temperatures the lowest group of modes is occupied, the converse is true for negative temperatures. When $T \to \pm\infty$, equipartition of power takes place among modes. **c**, Energy-temperature diagrams pertaining to this family of lattices corresponding to the same power levels in (**a** and **b**), shown as colored curves. In all cases, the number of modes was taken to be $M = 100$.

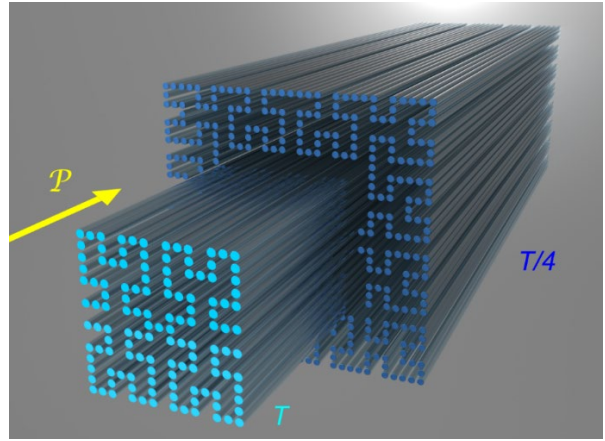

**Fig. 3 | Joule expansion in a nonlinear photonic multimoded chain system.** Light conveying power $\mathcal{P}$ is allowed to suddenly expand in a larger chain network having four times as many waveguide elements ($M \to 4M$). In this case, if the temperature of the photon gas before this transition is $T$, after this abrupt and thermodynamically irreversible expansion, the system reaches equilibrium at $T/4$ – a direct consequence of the way Joule expansion manifests itself in these heavily multimoded nonlinear optical environments.

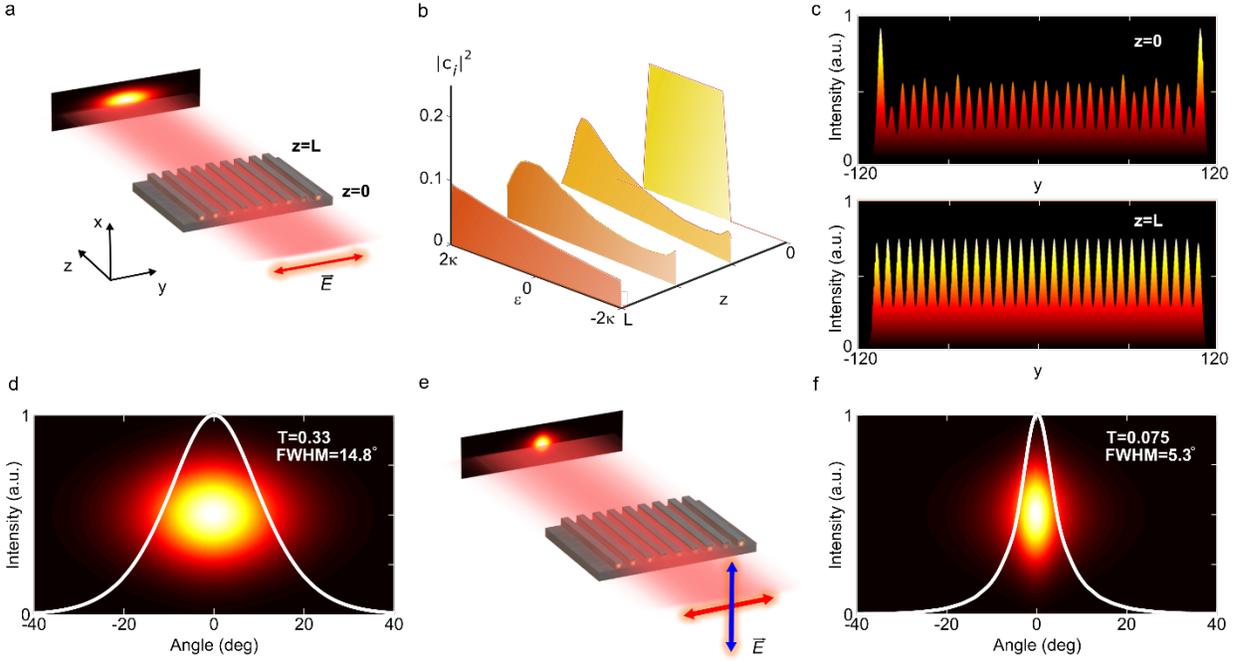

**Fig. 4 | Thermodynamic beam cleaning and cooling. a**, Propagation of a $\hat{y}$ polarized optical beam in a nonlinear chain of waveguides involving 30 elements. The projected far-field after the beam exits the system at $z = L$ is also shown. **b**, Thermalization dynamics of the modal occupancies as a function of distance when $\mathcal{P} = 2$ and the modes are uniformly excited in the range of $-0.5\kappa \leq \varepsilon_j \leq 1.2\kappa$. At the output, the system settles into a Rayleigh-Jeans distribution with a temperature of $T = 0.33$. **c**, The intensity distribution corresponding to the input used in (**b**) displays a strong speckle (upper panel). The beam self-cleans after thermalization takes place (lower panel). **d**, Far-field pattern associated with the thermalized $\hat{y}$ polarized output field for the same conditions used in (**b**). **e**, Cooling the $\hat{y}$ polarized wavefront using cross-phase modulation with an $\hat{x}$ polarized beam of equal power. **f**, After cooling, the far-field of the $\hat{y}$ polarized wave experiences considerable improvement. This cooling interaction can be analytically predicted[24] from Eqs. 2.

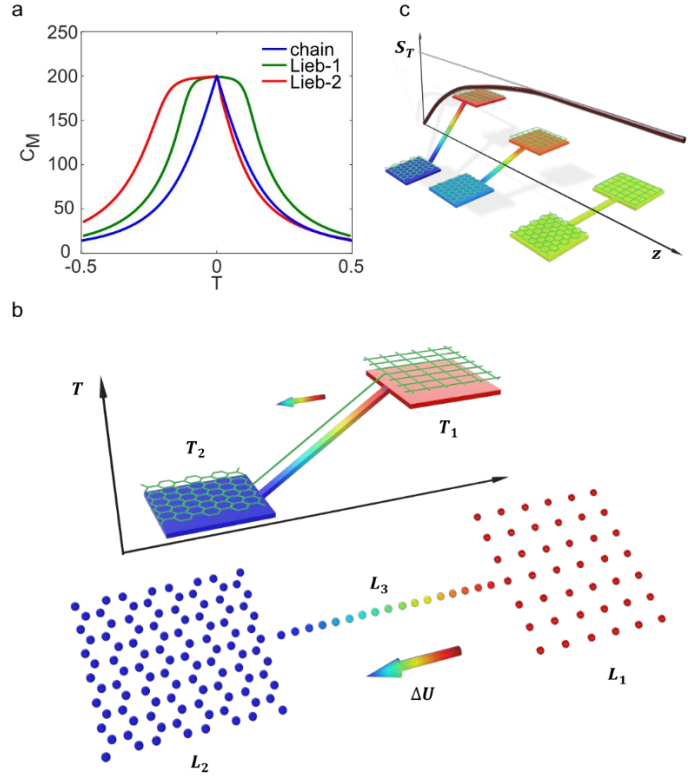

**Fig. 5 | Photo-capacity and non-equilibrium energy "heat" transfer. a**, Photo-capacity curves corresponding to a standard (Fig. 1) chain network (blue curve), a Lieb lattice (Lieb-1) with only nearest-neighbor interactions (green) and a Lieb array (Lieb-2) with cross-couplings (red), when $\mathcal{P} = 20$ and $M = 200$. Note the asymmetry in the photo-capacity curve of the Lieb-2 system. **b**, Energy transfer $\Delta U$ between a square $L_1$ and a graphene $L_2$ lattice when initially kept at different temperatures ($T_1 = 0.3, T_2 = 0.1$). This energy transfer takes place through a bridge chain composed of birefringent elements under non-equilibrium conditions. As required by the second law of thermodynamics, $\Delta U$ always flows from a hot to a cold subsystem. In this case, the temperature in the linking chain drops linearly, in accord with Fick's law. **c**, The internal energy transfer ceases when thermal equilibrium is reached in which case both subsystems attain the same temperature and the total entropy $S_T$ is maximized. In all cases, the nonlinearity was assumed to be of the Kerr type.

# Supplementary Methods

## Global equation of state

Using the expressions for the two conserved quantities associated with the optical power $\mathcal{P}$ and internal energy $U$, one can write

$$U - \mu \mathcal{P} = \sum_{j}^{M} \left[ -\varepsilon_j |c_j|^2 - \mu |c_j|^2 \right]$$

A direct substitution of the Rayleigh-Jeans distribution, $|c_j|^2 = -T/(\varepsilon_j + \mu)$ [24-26] into the above expression, leads to a global equation of state:

$$U - \mu \mathcal{P} = T \sum_{j}^{M} \frac{\varepsilon_j + \mu}{\varepsilon_j + \mu} = MT.$$

## $U - T$ relation associated with a multimode network chain of nonlinear coupled elements

Let us consider a 1D chain of weakly nonlinear elements. These elements are evanescently coupled via nearest neighbor interactions with an exchange strength $\kappa$. Starting from the Rayleigh-Jeans distribution, the power conservation law in a nonlinear multimode optical system can be written as:

$$\mathcal{P} = \sum_{j=1}^{M} |c_j|^2 = \sum_{j=1}^{M} -\frac{T}{\varepsilon_j + \mu}$$

On the other hand, the global equation of state requires that Given that the eigenspectrum of a 1-D discrete lattice is given by $\varepsilon_j = 2\kappa \cos\left(\frac{j\pi}{M+1}\right)$, we obtain

$$\sum_{j=1}^{M} \frac{1}{\frac{2\kappa \cos\left(\frac{j\pi}{M+1}\right)}{T} + \frac{1}{\mathcal{P}}\left(\frac{U}{T} - M\right)} = -\mathcal{P}$$

We now set $\beta = \frac{1}{T}, x = \frac{j\pi}{M+1}, dx = \frac{\pi}{M+1}$ and $A = \frac{2\beta\kappa\mathcal{P}}{\beta U - M} = \frac{2\kappa}{\mu} = \frac{2\kappa\mathcal{P}}{U - MT}$. Note that one can rigorously show that for positive temperatures $T > 0$, $MT > U + 2\kappa\mathcal{P}$ whereas for negative temperatures ($T < 0$) the following is true $MT < U - 2\kappa\mathcal{P}$. As a result, $-1 < A < 1$ or $A^2 < 1$. Since these structures support a large number of modes $M$, the summation above can now be approximated with an integral.

$$\Rightarrow \frac{(M+1)\mathcal{P}}{\pi(M - \beta U)} \int_{0}^{\pi} \frac{dx}{1 + A\cos(x)} = \mathcal{P}$$

$$\Rightarrow \frac{(M+1)\mathcal{P}}{\pi(M-\beta U)} \frac{2}{\sqrt{1-A^2}} \tan^{-1}\left[\frac{(1-A)\tan\left(\frac{x}{2}\right)}{\sqrt{1-A^2}}\right]\bigg|_0^\pi = \mathcal{P}$$

After applying the limits $\tan\left(\frac{\pi}{2}\right) = +\infty$, $\tan^{-1}(+\infty) = \frac{\pi}{2}$, we find

$$\Rightarrow \frac{(M+1)}{(M-\beta U)\sqrt{1-A^2}} = 1$$

Because $M$ is large, $M+1 \to M$, and given that $A = \frac{2\beta\kappa\mathcal{P}}{\beta U - M}, \beta = \frac{1}{T}$.

$$\Rightarrow \frac{M}{\sqrt{(M-\beta U)^2 - 4\beta^2\kappa^2\mathcal{P}^2}} = 1$$

$$\Rightarrow T = \frac{U^2 - 4\kappa^2\mathcal{P}^2}{2UM} \tag{S1}$$

$$\Rightarrow U = \begin{cases} TM - \sqrt{T^2M^2 + 4\kappa^2\mathcal{P}^2} & (T > 0) \\ TM + \sqrt{T^2M^2 + 4\kappa^2\mathcal{P}^2} & (T < 0) \end{cases}$$

The last three expressions in this section provide the $U - T$ relation associated with a multimode network chain of nonlinear coupled elements.

**Optical Sackur-Tetrode equation**

From the fundamental equation of thermodynamics, the temperature can be obtained from the entropy according to

$$\frac{\partial S}{\partial U} = \frac{1}{T}$$

Direct integration leads to

$$S(U, \mathcal{P}, M) = S(0, \mathcal{P}, M) + \int_0^U \frac{1}{T} du$$

After substituting Eq. (S1) into the last integral and by setting $B^2 = 4\kappa^2\mathcal{P}^2$

$$S = S(0, \mathcal{P}, M) + \int_0^U \frac{2uM}{u^2 - B^2} du$$

Since $U^2 < 4\kappa^2\mathcal{P}^2 = B^2$, the integral gives

$$S = S(0, \mathcal{P}, M) + M \ln\left(\frac{B^2 - U^2}{B^2}\right) = S(0, \mathcal{P}, M) + M \ln\left(\frac{4\kappa^2\mathcal{P}^2 - U^2}{4\kappa^2\mathcal{P}^2}\right)$$

According to Eq. (S1), when $U \to 0$, $T \to \infty$. In this limit, the Rayleigh-Jeans distribution implies that the mode occupancies are equipartitioned $|c_j|^2 \to -\frac{T}{\mu}$ and from the first global equation of state $U - \mu \mathcal{P} = MT \Rightarrow -\frac{T}{\mu} = \frac{\mathcal{P}}{M}$. Therefore

$$S(0, \mathcal{P}, M) = \lim_{U \to 0} \sum_j^M \ln\left(|c_j|^2\right) = M \ln\left(\frac{\mathcal{P}}{M}\right)$$

The total relative optical Sackur-Tetrode entropy of this particular system can now be written as a function of the extensive variables $U, \mathcal{P}, M$, as in Eq. (1):

$$S(U, \mathcal{P}, M) = M \ln\left(\frac{4\kappa^2 \mathcal{P}^2 - U^2}{4M\kappa^2 \mathcal{P}}\right)$$

**Absolute entropy**

Equation (1) represents a relative entropy from where the equations of state can be obtained. The absolute entropy $S_n$ can be obtained from first principles and involves an additional term that results from the minimum power/energy levels used in "quantizing" this statistical photo-mechanics problem. This quantization leads to an additional term $M \ln X$ in the absolute entropy, i.e.,

$$S_n = S + M \ln X$$

where $S = \sum_j^M \ln|c_j|^2$ and $X$ is a constant. Here, $M \ln X$ represents an entropy floor that keeps the entropy always positive. This last term plays a crucial role in accounting for the entropy increase during Joule expansion. On the other hand, all the thermodynamic relations derived here can simply obtained from the relative entropy $S$ without any loss of generality.

**Isentropic processes**

During an isentropic process, the waveguide/cavity array is adiabatically compressed or expanded so that the tight-binding coupling coefficient $\kappa$, will vary accordingly during evolution while the quantities $\mathcal{P}$ and $M$ in Eq. (1) remain constant during propagation. Because of adiabaticity, the modal occupancies remain invariant, and so does the entropy.

In order to keep the entropy $S$ in Eq. (1) constant, the ratio $U/\kappa$ must be constant, i.e. $U/\kappa = const$. From Eqs. (2a-b) of the main text, one can then easily deduce that $U/T = const$. and $\mu/T = const$.

An alternative approach to this problem is given by the following argument.

The partial derivative of the entropy $S$ with respect to $\kappa$ while $\mathcal{P}$ and $M$ are constant is given by

$$\left.\frac{\partial S}{\partial \kappa}\right|_{\mathcal{P},M} = \left.\frac{\partial S}{\partial \kappa}\right|_{U,\mathcal{P},M} + \left.\frac{\partial S}{\partial U}\right|_{\mathcal{P},M,\kappa} \left.\frac{\partial U}{\partial \kappa}\right|_{S,\mathcal{P},M} \tag{S2}$$

From the Sackur-Tetrode equation (1) and after using Eq. (S1) one finds that:

$$\left.\frac{\partial S}{\partial \kappa}\right|_{U,\mathcal{P},M} = \left(\frac{8M\kappa\mathcal{P}^2}{4\kappa^2\mathcal{P}^2 - U^2} - \frac{8M^2\kappa\mathcal{P}}{4M\kappa^2\mathcal{P}}\right)$$

$$= \frac{8M\kappa\mathcal{P}^2}{4\kappa^2\mathcal{P}^2 - U^2} - \frac{2M}{\kappa} = -\frac{U}{\kappa T}$$

In other words,

$$\left.\frac{\partial S}{\partial \kappa}\right|_{U,\mathcal{P},M} = -\frac{U}{\kappa T}$$

The optical Sackur-Tetrode equation can be re-written so as to express $U$ in terms of the $S, \mathcal{P}, M, \kappa$ variables in the following fashion:

$$U^2(S,\mathcal{P},M,\kappa) = 4\kappa^2\mathcal{P}^2 - 4M\kappa^2\mathcal{P}e^{\left(\frac{S}{M}\right)}$$

Hence the partial derivative corresponding to $\kappa$ is given by:

$$\left.\frac{\partial U}{\partial \kappa}\right|_{S,\mathcal{P},M} = \frac{\kappa}{U}\left(4\mathcal{P}^2 - 4M\mathcal{P}e^{\left(\frac{S}{M}\right)}\right)$$

After substituting this last result in Eq. (S2) and by keeping in mind that

$$\left.\frac{\partial S}{\partial U}\right|_{\mathcal{P},M,\kappa} = \frac{1}{T}$$

we find

$$\left.\frac{\partial S}{\partial \kappa}\right|_{\mathcal{P},M} = -\frac{U}{\kappa T} + \frac{1}{T}\frac{\kappa}{U}\left(4\mathcal{P}^2 - 4M\mathcal{P}e^{\left(\frac{S}{M}\right)}\right) = 0$$

In other words the entropy remains constant when the coupling strength of the array adiabatically changes during evolution. In addition, the isentropic process demands that

$$dS|_{\mathcal{P},M} = \left.\frac{\partial S}{\partial \kappa}\right|_{U,\mathcal{P},M} d\kappa + \left.\frac{\partial S}{\partial U}\right|_{\mathcal{P},M,\kappa} dU = -\frac{U}{\kappa T}d\kappa + \frac{1}{T}dU = 0$$

$$\Rightarrow \frac{d\kappa}{\kappa} = \frac{dU}{U}$$

$$\Rightarrow \frac{\kappa}{U} = const.$$

Hence $\frac{U}{T} = const.$ and $\frac{\mu}{T} = const.$ as found before.

**Equations of state**

From Eq. (1) we obtain:

$$S = M \ln(4\kappa^2 \mathcal{P}^2 - U^2) - M \ln(4M\kappa^2 \mathcal{P})$$

The equations of state can be derived after taking the partial derivatives with respect to the corresponding extensive variables:

$$\frac{\partial S}{\partial U} = \frac{-2MU}{4\kappa^2 \mathcal{P}^2 - U^2} = \frac{1}{T}$$

$$\Rightarrow T(U, \mathcal{P}, M) = \frac{U^2 - 4\kappa^2 \mathcal{P}^2}{2UM} \tag{S3}$$

$$\frac{\partial S}{\partial \mathcal{P}} = \frac{8M\kappa^2 \mathcal{P}}{4\kappa^2 \mathcal{P}^2 - U^2} - \frac{4M^2\kappa^2}{4M\kappa^2 \mathcal{P}} = -\alpha$$

$$\Rightarrow \alpha(U, \mathcal{P}, M) = \frac{\mu}{T} = \frac{M}{\mathcal{P}} + \frac{8M\kappa^2 \mathcal{P}}{U^2 - 4\kappa^2 \mathcal{P}^2} \tag{S4}$$

$$\frac{\partial S}{\partial M} = \ln(4\kappa^2 \mathcal{P}^2 - U^2) - \ln(4M\kappa^2 \mathcal{P}) - \frac{4M\kappa^2 \mathcal{P}}{4M\kappa^2 \mathcal{P}} = \frac{\hat{p}}{T}$$

$$\Rightarrow \frac{\hat{p}}{T} = \ln\left(\frac{4\kappa^2 \mathcal{P}^2 - U^2}{4M\kappa^2 \mathcal{P}}\right) - 1 \tag{S5}$$

The global equation of state is equivalent to the above set of equations (S3-S5). According to Eq. (S3)

$$4\kappa^2 \mathcal{P}^2 = U^2 - 2UMT \tag{S6}$$

Substitution of Eq. (S6) into Eq. (S4) so as to eliminate the term $4\kappa^2 \mathcal{P}^2$, leads to the global equation of state:

$$\Rightarrow U - \mu \mathcal{P} = MT$$

In addition, the Euler equation is also equivalent to the above three equations of state since:

$$\frac{U}{T} - \frac{\mu}{T}\mathcal{P} + \frac{\hat{p}}{T}M$$
$$= \frac{2U^2 M}{U^2 - 4\kappa^2 \mathcal{P}^2} - M - \frac{8M\kappa^2 \mathcal{P}^2}{U^2 - 4\kappa^2 \mathcal{P}^2} + M \ln\left(\frac{4\kappa^2 \mathcal{P}^2 - U^2}{4M\kappa^2 \mathcal{P}}\right) - M$$
$$= \frac{2M(U^2 - 4\kappa^2 \mathcal{P}^2)}{U^2 - 4\kappa^2 \mathcal{P}^2} - 2M + M \ln\left(\frac{4\kappa^2 \mathcal{P}^2 - U^2}{4M\kappa^2 \mathcal{P}}\right)$$
$$= M \ln\left(\frac{4\kappa^2 \mathcal{P}^2 - U^2}{4M\kappa^2 \mathcal{P}}\right) = S$$

**Final temperature of the optical cooling process**

The final temperature of any cooling process involving two identical chain subsystems (each conserving its power and having the same number of modes $M$) can be uniquely determined from the initial power $\mathcal{P}_1, \mathcal{P}_2$ and energy $U_{10}, U_{20}$ by using Eq. (2a). Since eventually, at thermal equilibrium, they share the same temperature $T$ while exchanging energy, this temperature can be obtained from Eq. (2a)

$$\frac{U_1^2 - 4\kappa_1^2 \mathcal{P}_1^2}{2MU_1} = T = \frac{U_2^2 - 4\kappa_2^2 \mathcal{P}_2^2}{2MU_2}$$

where $U_1$ and $U_2$ represent the "final energy" in each system. In the above equation, $\kappa_1$ and $\kappa_2$ represent the coupling constants corresponding to the two different polarizations in Fig. 4. Given that the total energy $U_T = U_1 + U_2 = U_{10} + U_{20}$ is conserved, the above equation leads to a cubic equation for $U_1$:

$$2U_1^3 - 3U_T U_1^2 + (U_T^2 - 4\kappa_1^2 \mathcal{P}_1^2 - 4\kappa_2^2 \mathcal{P}_2^2)U_1 + 4\kappa_1^2 \mathcal{P}_1^2 U_T = 0$$

Once $U_1$ is obtained from the above equation (of the three possible roots, the physical solution should be in the range of $-2\kappa \mathcal{P}_1 \leq U_1 \leq 2\kappa \mathcal{P}_1, -2\kappa \mathcal{P}_2 \leq U_2 \leq 2\kappa \mathcal{P}_2$), the final temperature can be determined from

$$T = \frac{U_1^2 - 4\kappa^2 \mathcal{P}_1^2}{2MU_1}.$$

Finally the chemical potential in every subsystem can be extracted from the global equation of state ($U - \mu \mathcal{P} = MT$). From here, the Rayleigh-Jeans distribution in each subsystem follows.

**Normalizations used in multimode nonlinear lattice systems**

In a 1-D discrete nonlinear photonic arrangement, the evolution of the optical modal complex amplitude $A_m$ at the $m^{th}$ site is governed by:

$$i\frac{dA_m}{dZ} + \beta_m A_m + \eta(A_{m+1} + A_{m-1}) + n_2 k_0 |A_m|^2 A_m = 0,$$

where $\beta_m$ is the propagation constant at the corresponding site, $\eta$ is the nearest-neighbor hopping coefficient, $k_0$ is the wavenumber in vacuum and $n_2$ represents the effective nonlinear Kerr coefficient. When the photonic lattice is truncated to have $M$ sites, the boundary condition is $a_0 = a_{M+1} = 0$. By assuming that all sites are identical ($\beta_m = \beta_0$), and by setting $z = \eta_0 Z, \kappa = \eta/\eta_0$, $a_m = A_m e^{-i\beta_0 Z}/\rho, \gamma = n_2 k_0 \rho^2/\eta_0 = 1$, one obtains the normalized evolution equations:

$$i\frac{da_m}{dz} + \kappa(a_{m+1} + a_{m-1}) + |a_m|^2 a_m = 0.$$

The above equation of motion (in the local mode representation) can be derived from the system total Hamiltonian $H_T$ that is composed by a linear and a nonlinear part:

$$H_T = H_L + H_{NL} = \sum_{m=1}^{M} \kappa(a_{m+1} a_m^* + a_{m-1} a_m^*) + \sum_{m=1}^{M} |a_m|^4 \qquad (S7)$$

The linear part of Hamiltonian can be readily written in the supermode representation by substituting $|\psi\rangle = \sum_{j=1}^{M} c_j(\xi)|\phi_j\rangle$ into Eq. (S7), i.e.,

$$H_L = \sum_{j=1}^{M} \varepsilon_j |c_j|^2$$

The arrays are supposed to be lossless and having no leakage to radiation modes. The narrow band structure formed by the fundamental mode is expected to lie away from the radiation modes of the system.

**Lieb lattice**

Each unit cell in the Lieb lattice contains three sites, indicated by red, green and blue. When only the vertical and horizontal nearest neighbour couplings ($\kappa_1 = 1$) are taken into account, the Lieb lattice (Lieb-1) displays a highly degenerate flat band at the centre of the eigen-spectrum. When the cross-coupling ($\kappa_2 = 0.5$) is present, the band structure of the Lieb lattice becomes asymmetric (Lieb-2).

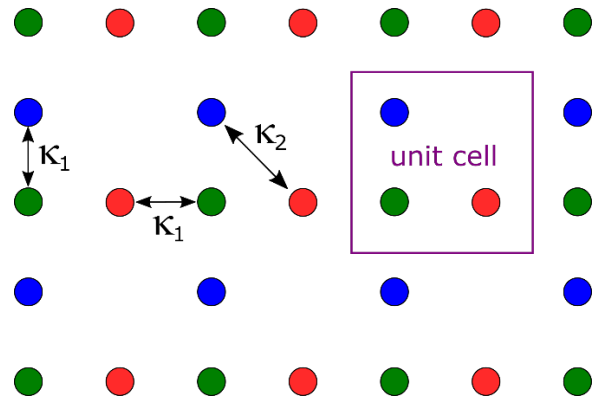

**FIG. S1** | Geometric structure of a Lieb lattice.

## Hexagonal, chain and square lattices

The exchange coupling mechanisms between nearest neighbors in discrete hexagonal, chain and square lattices are presented schematically in Fig. S2.

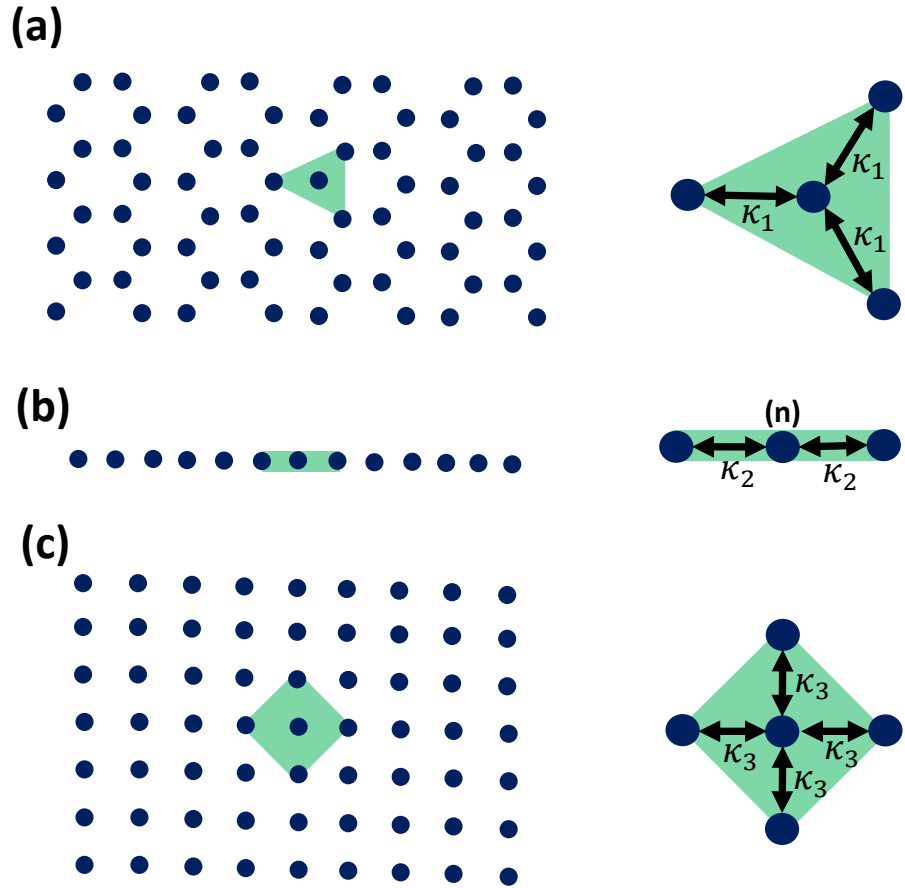

**FIG. S2** | Coupling exchange $\kappa$ in a (a) hexagonal, (b) optical chain, (c) square lattice. In each arrangement, only interactions between nearest-neighbors are involved.

**Governing nonlinear equations for the two orthogonal polarizations**

In a chain network, the discrete coupled nonlinear Schrödinger equations describing the evolution of the two orthogonal polarizations $a_n$ and $b_n$ associated with the normalized optical field at site $n$ are given by:

$$i\frac{da_n}{dz} + \beta_n^a a_n + \kappa_a(a_{n+1}+a_{n-1}) + A|a_n|^2 a_n + B|b_n|^2 a_n + C(b_n)^2 a_n^* = 0 \quad (S8)$$

$$i\frac{db_n}{dz} + \beta_n^b b_n + \kappa_b(b_{n+1}+b_{n-1}) + A|b_n|^2 b_n + B|a_n|^2 b_n + C(a_n)^2 b_n^* = 0 \quad (S9)$$

Here $n$ denotes the site number, $\kappa_{a(b)}$ the coupling coefficients, $\beta_n^{a,b}$ the propagation constants of these waveguides, and the last three terms correspond to self-phase modulation, cross-phase modulation and four-wave mixing effects. The nonlinear coefficients $A$, $B$ and $C$ result from the $\chi^{(3)}$ tensor. If the waveguide birefringence $(\beta^a - \beta^b)$ in every element is strong with respect to the nonlinearly induced index change, the four-wave mixing process can be ignored ($C = 0$) in the above equations. Throughout this study we have used $A = 1, B = 2/3$ and $C = 0$., values corresponding to silica glass. In the same vein, the governing equations for the two optical polarizations can be readily written for hexagonal and square lattices.

## Thermal conductivity simulations

The simulations associated with thermal conductivity were performed for an arrangement where a photonic square lattice $L_1$ with $M_1$ sites is connected to a graphene-like system $L_2$ with $M_2$ sites via a chain bridge $L_3$ involving $M_3$ evanescently coupled waveguides. The chain bridge consists of 7 sub-chains $(L_3^{(1)}, L_3^{(2)}, \ldots, L_3^{(7)})$ with $M_3^{(s)}$ number of sites (see Fig. S3). Each optical array conveys a different polarization (blue and red sides represents $\hat{x}$ and $\hat{y}$ polarizations). On the other hand, the polarizations are allowed to overlap in the thermal bridge layer (purple) in between. This bridge "diathermal" layer allows only for energy $U$ transfer from one to the other optical array via cross phase modulation. In such arrangements the field evolution is governed by Eqs. S8-S9.

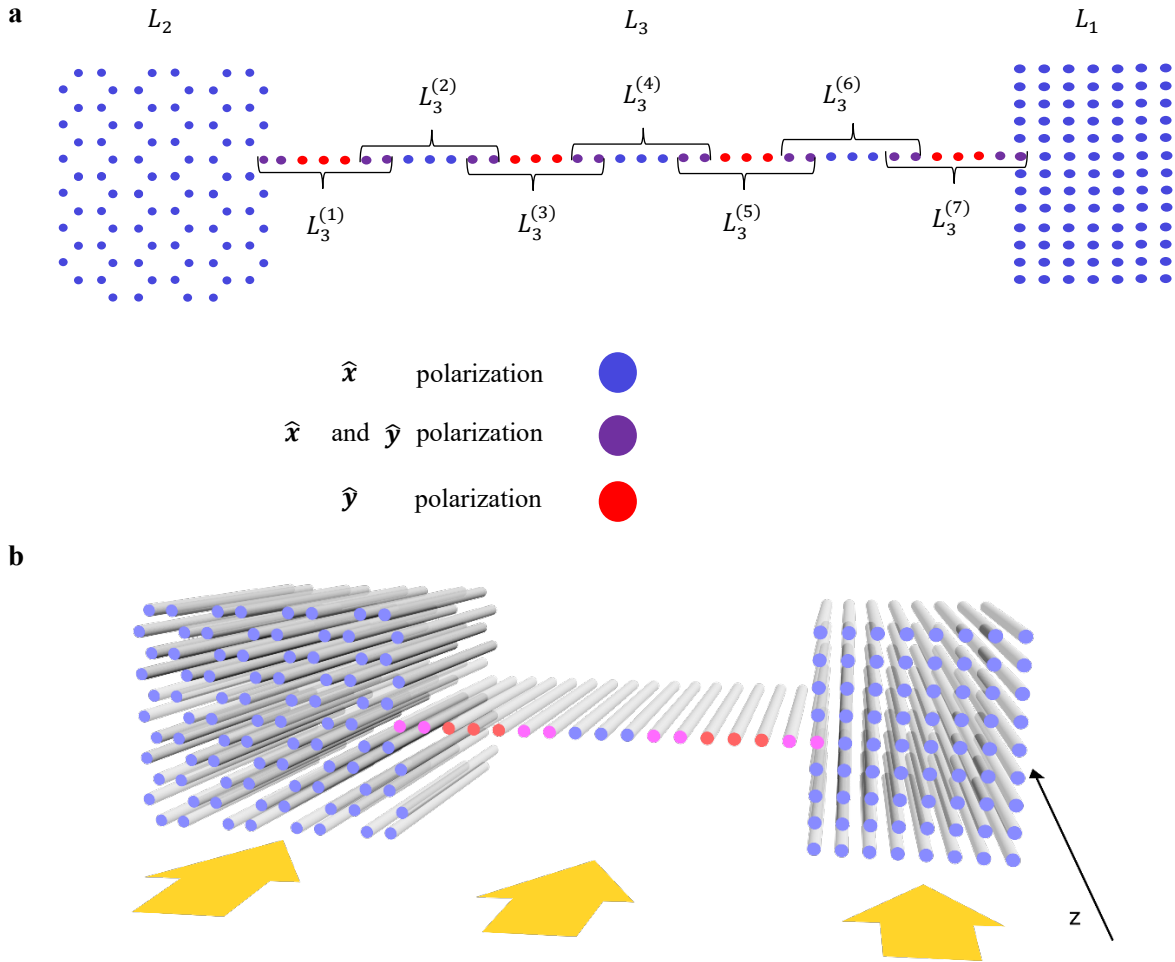

**Fig. S3 | a**, The arrangement used to simulate energy $U$ transport phenomena occurring when a hot (square lattice) and a cold (hexagonal array) optical nonlinear multimode systems are linked through an optical chain. In the simulations $M_1 = 900$, $M_2 = 896$, $M_3 = 140$ and $M_3^{(s)} = 20$. At every junction (purple) in the thermal bridge, two sites are assigned between nearest optical arrays where cross-phase modulation effects take place. **b**, A schematic illustration of the photonic lattices described in (**a**)

## Chain design

A possible integrated optical chain design on $Si_3N_4/SiO_2$ platform is shown below for $\lambda = 1.064 \mu m$. The governing equations are of the type indicated in Eqs. S8-S9.

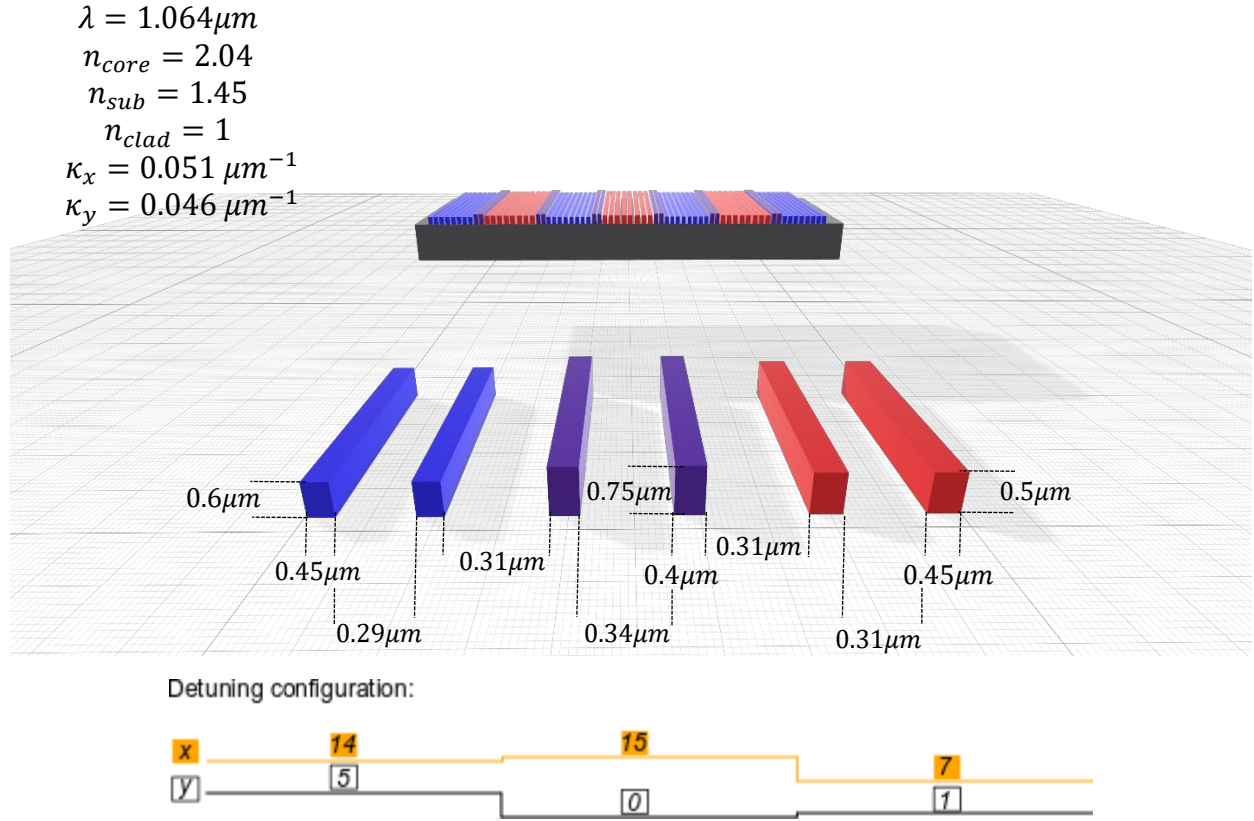

**Fig. S4** | Possible design for the chain bridge (thermal link). The design involves three groups of birefringent sites, based on different geometries. At the blue and red sites, the $\hat{x}$ and $\hat{y}$ polarizations are confined, respectively. Purple sites represent "thermal layers" where the two polarizations are allowed to overlap and interact via nonlinear cross-phase modulation. This can be achieved by tuning the propagation constants for the $\hat{x}$ and $\hat{y}$ polarizations (see yellow and black lines below the figure). In such an optical setting, the left (blue) and right (red) groups act as the two subsystems while the middle group plays the role of a diathermal layer.